%% file: main.tex
\documentclass[conference,a4paper]{IEEEtran}
\usepackage{graphicx,xcolor,tikz}
\usepackage{comment}
\usepackage{subfig}
\usepackage{float}
\usepackage{multicol}
\usepackage{cite}
\usepackage{makecell}
\usepackage{siunitx}
\usepackage{multirow}
\usepackage{mathtools}
\usepackage{amsmath,amssymb,amsmath,amsfonts}
\usepackage{cite}
\usepackage{caption}
\usepackage{lipsum}

\usepackage{epsfig,graphics,graphicx,subfig,amssymb,amstext,amsmath,algorithm,algorithmic,wrapfig,multirow}
\usepackage{adjustbox}
\usepackage{pgfplots}
\usepackage{xcolor}
\usetikzlibrary{chains,arrows,calc ,positioning}
\usepackage{graphicx}
\usepackage{colortbl, xcolor}
\usepackage{enumitem}
\setitemize{noitemsep,topsep=0pt,parsep=0pt,partopsep=0pt}
\definecolor{maroon}{cmyk}{0,0.87,0.68,0.32}

\pgfplotsset{compat=1.16}
\definecolor{bittersweet}{rgb}{1.0, 0.44, 0.37}
\definecolor{glaucous}{rgb}{0.38, 0.51, 0.71}
\definecolor{gainsboro}{rgb}{0.86, 0.86, 0.86}
\definecolor{babyblueeyes}{rgb}{0.63, 0.79, 0.95}
\definecolor{silver}{rgb}{0.75, 0.75, 0.75}
\definecolor{neoncarrot}{rgb}{1.0, 0.64, 0.26}

\usepackage[normalem]{ulem}

\def\ISDs   {{\mathsf{ISD}_{\sf s}}}

\makeatletter

\providecommand{\tabularnewline}{\\}

 \let\oldforeign@language\foreign@language
 \DeclareRobustCommand{\foreign@language}[1]{%
   \lowercase{\oldforeign@language{#1}}}

\usepackage[font=small]{caption}

\usepackage{dblfloatfix}

\input{commands.tex}

\makeatother

\IEEEoverridecommandlockouts

\begin{document}

\bstctlcite{IEEEexample:BSTcontrol}

%
\title{Coexistence of UAVs and Terrestrial Users in Millimeter-Wave Urban Networks}
%

\author{
%
\IEEEauthorblockN{Seongjoon Kang$^{\dagger}$ \quad Marco Mezzavilla$^{\dagger}$ \quad Angel Lozano$^{\flat}$  \quad Giovanni Geraci$^{\flat}$ \\ \quad Sundeep Rangan$^{\dagger}$   \quad Vasilii Semkin$^{\sharp}$ \quad William Xia$^{\dagger}$ \quad Giuseppe Loianno$^{\dagger}$ \vspace{0.2cm}} 
\IEEEauthorblockA{$^{\dagger}$NYU Tandon School of Engineering, Brooklyn, NY, USA}
\IEEEauthorblockA{$^{\flat}$Univ. Pompeu Fabra, Barcelona, Spain}
\IEEEauthorblockA{$^{\sharp}$VTT Technical Research Centre of Finland Ltd, Finland}
\thanks{S. Rangan, W. Xia, S. Kang, and M. Mezzavilla were supported by NSF grants  1302336,  1564142,  1547332, and 1824434, SRC, and the industrial affiliates of NYU WIRELESS. A.~Lozano and G. Geraci were supported by ERC grant 694974, by MINECO's Projects RTI2018-101040 and PID2021-123999OB-I00, by the ``Ram\'{o}n y Cajal" program, and by ICREA. The work of V. Semkin was supported in part by the Academy of Finland.}
}

\maketitle

\begin{abstract}
\input{00_abstract.tex}
\end{abstract}


\IEEEpeerreviewmaketitle

\input{01_intro.tex}
\input{02_system_model.tex}

\input{03_standard_cells.tex}
\input{04_dedicated_cells.tex}
\input{05_conclusion.tex}

\bibliographystyle{IEEEtran}
\bibliography{bibl.bib}

\end{document}

%% file: commands.tex
\def\nb0{{\mathbf{0}}}
\def\nb1{{\mathbf{1}}}



%% file: 00_abstract.tex
5G millimeter-wave (mmWave) cellular networks are in the early phase of commercial deployments and present a unique opportunity for robust, high-data-rate communication to unmanned aerial vehicles (UAVs). 
A fundamental question is whether and how 
mmWave networks designed for terrestrial users should be modified to serve UAVs.
The paper invokes realistic cell layouts, antenna patterns, and channel models
trained from extensive ray tracing data to assess the performance of various
network alternatives. Importantly, the study considers the addition of dedicated uptilted rooftop-mounted cells for aerial coverage,
as well as novel spectrum sharing modes between terrestrial and aerial network operators. The effect of power control and of multiuser multiple-input multiple-output are also studied.


%% file: 01_intro.tex
\section{Introduction}

The interest is growing rapidly in unmanned
aerial vehicles (UAVs) for applications such as monitoring, surveying, precision agriculture, construction, remote sensing, or product delivery \cite{GerGarAza2022,ZenGuvZha2020,SaaBenMoz2020,zeng2019accessing,CheJaaYan2020}.
Low-latency connectivity at high data rates is required 
for many of these applications and the millimeter wave (mmWave) frequency range offers bandwidths that can be instrumental to meet these requirements 
\cite{xiao2016enabling,zhang2019research,mozaffari2019ControlsDroneAntenna,lee2018spectrum,shafi20175g}.
Also, links to UAVs are often line-of-sight (LOS), which is desirable due to the limited diffraction at these frequencies 
\cite{khawaja2017uav,heimann2018pencilbeam}. 

There has been extensive work in UAV wireless communication
and its interaction with cellular networks for frequencies below 6 GHz
 \cite{GerGarGal2018,lin2018sky,ChoGuvSaa2021,GarGerLop2019,BenGerLop2022,GerLopBen2022}.
Likewise, mmWave communication for terrestrial cellular networks has
seen enormous progress \cite{RanRapE:14,rappaportmillimeter,agiwal2016next}, 
particularly with the development
of the 3rd Generation Partnership Project (3GPP) 5G standard \cite{6824752}.
However, the coordination in the use of
spectrum between aerial and terrestrial networks in the mmWave realm faces
unique challenges. Chiefly, due to the directional nature of mmWave
transmissions \cite{roh2014millimeter}, 
the interference between aerial and terrestrial links is intricate, particularly in dense urban
scenarios with substantial ground and building reflections. 

In this paper, we investigate the coexistence of UAVs and terrestrial users (UEs) at mmWave frequencies under different spectrum sharing paradigms, thus bringing to the next level previous work that only focused on coverage~\cite{sjmma}. To this end, we conduct extensive system-level simulations with 3GPP channel models and antenna patterns for terrestrial users~\cite{3GPP38901,3GPP37840}, and with data-driven aerial channel models along with realistic antenna patterns for UAVs~\cite{XiaRanMez2020}.

We begin by considering a \emph{standard} mmWave deployment, with a single operator providing connectivity to both UAVs and UEs, and investigate the UAV-UE interplay under single- (SU) and multiuser (MU) multiple-input multiple output (MIMO) operation. This first study shows that:
\begin{itemize}
\item With SU-MIMO, UAV-to-UE interference is rather negligible because (i) urban LOS probability increases only marginally with the UAV height, (ii) the arrays of standard base stations (BSs) are downtilted, and (iii) highly directional mmWave transmissions provide considerable cell isolation.
\item With MU-MIMO, the data rates of both UAVs and UEs improve substantially as the interference remains low while multiple users are spatially multiplexed. 
\end{itemize}
We further study a setup with two mobile network operators (MNOs) sharing the same spectrum: a terrestrial MNO running a standard mmWave network and an aerial MNO operating \emph{dedicated} rooftop-mounted cells exclusively for UAVs. For this multi-MNO setup, we consider both \emph{closed access} and \emph{open access} paradigms, respectively confining UAVs to the aerial MNO and allowing UAVs to connect to whichever MNO offers the best service. This second study reveals that:
\begin{itemize}
\item Ensuring UAV coverage under closed access requires high densities of dedicated cells, yet this does not significantly penalize UEs; the increased number of concurrent UAV interferers is balanced by the fact that UAVs can afford transmitting at lower power.
\item Under open access, dedicated cells are not crucial for coverage, but their addition improves the performance of both UEs and UAVs. Indeed, UEs can now access resources previously taken up by UAVs, and UAVs enjoy shorter links and better antenna gains.
\end{itemize}

\begin{figure}[t!]
\centering
\includegraphics[width=1.0\columnwidth]{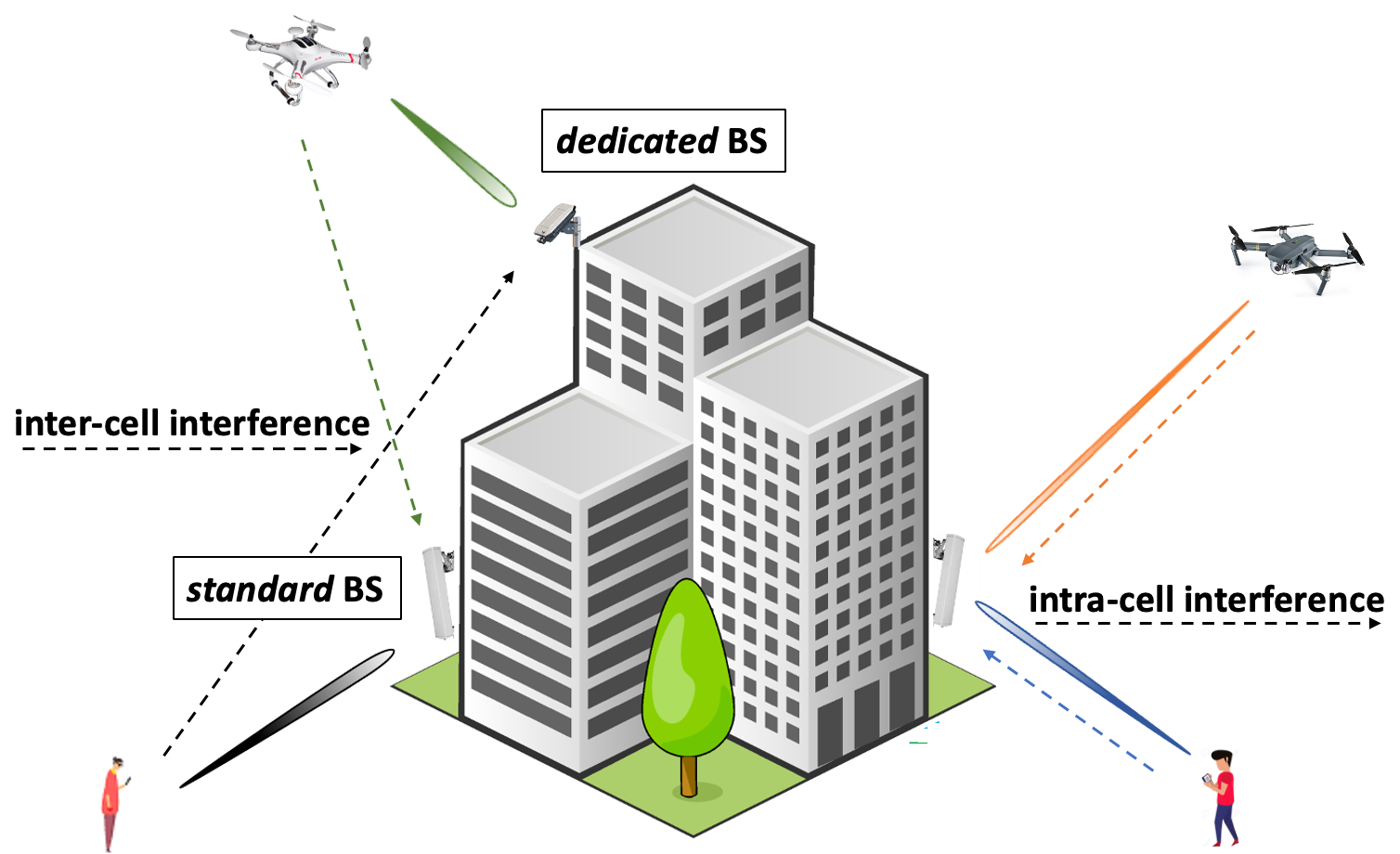}
\caption{Coexisting UAV and terrestrial UEs in a mmWave network.}
\label{fig:scenario}
\end{figure}

%% file: 02_system_model.tex
\section{System Model}
\label{sec:sim}

This work focuses on the uplink---the more data-hungry direction for UAVs---with the terrestrial UEs, UAVs, and BSs, all deployed in a $1 \, \text{km}^2 \times 1 \, \text{km}^2$ area with wrap-around.

\begin{table}[!t]
  \begin{center}
    \caption{Simulation parameters.}
    \label{tab:sim_params}
    \begin{tabular}{|l|l|}
    \hline
      \textbf{Parameter} & \textbf{Value}\\
      \hline
      Minimum UE-BS 2D distance (\SI{}{\meter}) &  $10$ \\
      Minimum UAV-BS 3D distance (\SI{}{\meter}) &  $10$ \\
      $\ISDs$ (\SI{}{\meter}) & $200$\\
      Height of standard BSs (\SI{}{\meter}) &  $10$\\ 
      Height of dedicated BSs (\SI{}{\meter}) & [10,30] \\
      Bandwidth (\SI{}{\MHz}) & $400 $ \\
      Frequency (\SI{}{\GHz}) & $28 $ \\
      BS noise figure (\SI{}{\dB}) & $6$\\
      UAV height (\SI{}{\meter}) & $120$ \\
      UAV and UE maximum transmit power (\SI{}{dBm}) & $23$\\
      3GPP vertical half-power beamwidth ($\theta_{\text{3dB}}$)  & $65^\circ$\\
      3GPP horizontal half-power beamwidth ($\phi_{\text{3dB}}$) & $65^\circ$ \\
      Power control parameters, $P_0$ (\SI{}{dBm}) and $\alpha$ & -$82$, $0.8$ \\
      Number of associated users per cell & $10$\\
      \hline
    \end{tabular}
  \end{center}
\end{table}

\subsection{Deployment and Propagation Channel Features}

\subsubsection*{Network Deployment} BSs are deployed following a homogeneous Poisson point process with a given average intersite distance (ISD), 
and they have three sectors, each corresponding to a cell and equipped with an $8 \times 8$ uniform rectangular array (URA). 
Two types of BSs are considered: a standard deployment with antennas downtilted by $-12^\circ$, and a dedicated deployment with rooftop-mounted antennas uptilted by $45^\circ$, as illustrated in Fig.~\ref{fig:scenario}.
In turn, UAVs and terrestrial users feature $4 \times 4$ URAs. Those mounted on UAVs are downtilted by $-90^\circ$ while those at terrestrial UEs are distributed uniformly in azimuth with a fixed elevation of $0^\circ$. All the parameters are summarized in Table \ref{tab:sim_params}, following the simulation settings in~\cite{3GPP38901}. The height of dedicated BSs, which is an additional parameter, is uniformly random between $10$ and $30$ m.

\subsubsection*{UAV Radiation Pattern} 
\begin{figure}[t!]
\centering
\includegraphics[width=0.7\columnwidth]{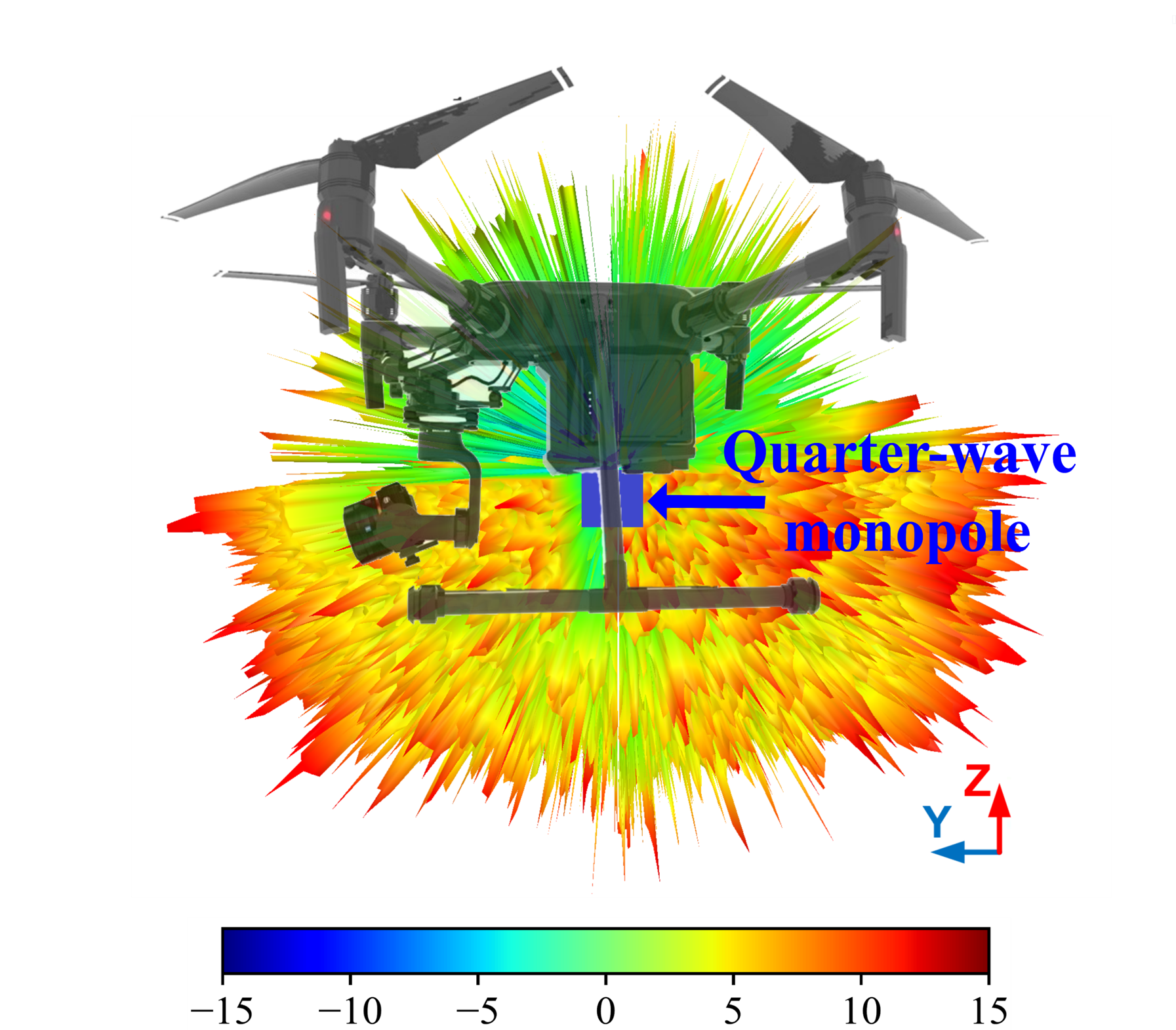}
\caption{Antenna radiation pattern accounting for the UAV frame.}
\label{fig:uav_antenna}
\vspace{-4mm}
\end{figure}

In order to incorporate the effects of the UAV frame on its radiation pattern, we simulated a quarter-wave monopole antenna (with bottom hemisphere coverage) placed under the UAV and used a software package~\cite{Remcom} that utilizes geometrical optics, uniform theory of diffraction, and physical optics. 
Fig.~\ref{fig:uav_antenna} illustrates the effect of the UAV on the antenna radiation pattern. The fluctuations of the antenna gain are due to protruding parts of the UAV (e.g., landing skid, gimbal) that are within the field of view of the antenna \cite{semkin2021lightweight}. Non-obviously, the effects of the UAV frame increase with the antenna directivity. We apply the obtained element radiation pattern to each UAV, whereas 3GPP antenna element patterns are adopted for BSs and UEs~\cite{3GPP37840} .

\subsubsection*{Propagation Model} The 3GPP channel model reported in~\cite{3GPP38901} is invoked to characterize the propagation between UEs and BSs (for urban microcells). To represent the mmWave propagation between UAVs and BSs, for which no calibrated model is available,
the data-driven channel model developed in \cite{XiaRanMez2020,XiaRanMez2020a} is invoked.

\subsection{Cell Selection and Beamforming}

Each transmitter, UE or UAV, connects to the strongest BS.
Both UAVs and UEs employ the open-loop power control policy specified in \cite{3GPP36213}. 
UAVs and UEs apply long-term transmit beamforming, aligning their beamforming vectors with the maximum-eigenvalue eigenvector of their channel covariance matrices~\cite{lozano2007long}.  After TX
beamforming, we obtain a SIMO channel 
$\boldsymbol{h}_{\ell k}$ and signal to noise ration (SNR) $\mathrm{SNR}_{\ell k}$ 
from user $k$ and BS $\ell$.
With MU-MIMO, the minimum mean-square error (MMSE) receive vector for user $u$ at BS $\ell$ is given by~\cite{heath2018foundations}
\begin{align}
  \boldsymbol{w}_{\ell u} & =  \Bigg(\sum_{k=0}^{N_{\rm u}-1} \mathrm{SNR}_{\ell k} \, \boldsymbol{h}_{\ell k} \boldsymbol{h}_{\ell k}^{*} \nonumber \\ 
  & \quad + \Bigg(1+\sum_{n \neq \ell } \sum_{k=0}^{N_{\rm n}-1} \mathrm{SNR}_{n k} \Bigg) \boldsymbol{I} \Bigg)^{\!-1} \!\! \boldsymbol{h}_{\ell u},
\end{align}
where the summation is only \emph{active} users scheduled in a particular time slot.
The out-of-cell interference is represented with the sum over $n\neq \ell$ where $N_n$
is the number of active users in cell $n$. 
As the channel matrices for other-cell users are not estimated by BS $\ell$, their interference is regarded as spatially white.
Accordingly, the overall gain from user $k$ to BS $\ell$ is
\begin{equation}
    G_{\ell k} = |\boldsymbol{w}_{\ell k}^{*} \boldsymbol{h}_{\ell k}|^2
\end{equation}
The interference experienced by user $u$ is then 
\begin{align}
    I_{\ell u} & = \sum_{k \neq u} P_{\ell k}^{\mathrm{tx}} G_{\ell k}
    \sum_m{A_{\ell k m}^{\rm \mathrm{tx}} A_{\ell k m}^{\rm rx} 
    L_{\ell k m}^{-1}} \nonumber \\
    &  \quad + \sum_{n \neq \ell} \sum_{k} P_{nk}^{\mathrm {tx}}G_{n k}  
    \sum_m A_{n k m}^{\mathrm {tx}} A_{nkm}^{\mathrm {rx}} L_{n k m}^{-1},
\end{align}
where $P_{\ell k}^{\mathrm{tx}}$ is the  transmit power of user $k$ in cell $\ell$, $A_{\ell km}^{\mathrm{tx}}$ and $A_{\ell km}^{\mathrm{rx}}$ are the transmit and receive antenna element gains on path $m$, and $L_{\ell km}$ is the pathloss along path $m$ between user $k$ and BS $\ell$. Similarly, the received power from user $u$ is
\begin{equation}
        P_{\ell u}^{\mathrm{rx}} = P_{\ell u}^{\mathrm{tx}} G_{\ell u}\sum_m{A_{\ell um}^{\mathrm{tx}} A_{\ell um}^{\mathrm{rx}} L_{\ell um}^{-1}}.
\end{equation}
Altogether, the SINR between UE $u$ and BS $\ell$ equals
\begin{align}
    \mathrm{SINR}_{\ell u} = \frac{P_{\ell u}^{\mathrm{rx}}}{I_{\ell u} +  N_0 B \,
    \|\boldsymbol{w}_{\ell u}\|^2},
\end{align}\\
where $N_0$ is the noise power spectral density and $B$ the bandwidth. 

%% file: 03_standard_cells.tex
\section{UAV-UE Spectrum Sharing with\\Standard mmWave Cells}
\label{sec:standard}

We begin by considering a standard mmWave deployment. This is handled by a single terrestrial MNO that provides connectivity to both UAVs and terrestrial UEs. For this setup, we investigate the coexistence and interplay between the two populations of users under SU-MIMO and MU-MIMO. At this stage, no dedicated BSs are assumed.


\subsection{SU-MIMO mmWave Communication}
 
To evaluate the amount of interference introduced by UAVs depending on their penetration rate and power control, we examine the three configurations reported in Table~\ref{tab:pwc_config}, namely with (1) no UAVs in the network, (2) a fraction of 50\% of UAVs employing fractional power control, and (3) a fraction of 50\% UAVs transmitting at maximum power. Since the network operates in SU-MIMO mode, each BS schedules a single user at random on a given time-frequency physical resource block. In configurations (2) and (3), half of the scheduled users on average are UAVs.

\begin{table}[h]
  \begin{center}
    \caption{UAV penetration rates and power control considered.}
    \label{tab:pwc_config}
    \begin{tabular}{|c|c|c|c|}
    \hline
      \textbf{Configuration} & \textbf{UAV fraction} & \textbf{UAV power control} \\
      \hline
        $1$ & {$0\%$} & None  \\
      \hline
      $2$ & {$50\%$} & Open loop  \\
      \hline
      $3$ & {$50\%$} & Max ($23$ \text{dBm})  \\
      \hline
    \end{tabular}
  \end{center}
\end{table}

\begin{figure}[t]
 \centering
 \includegraphics[width=\columnwidth]{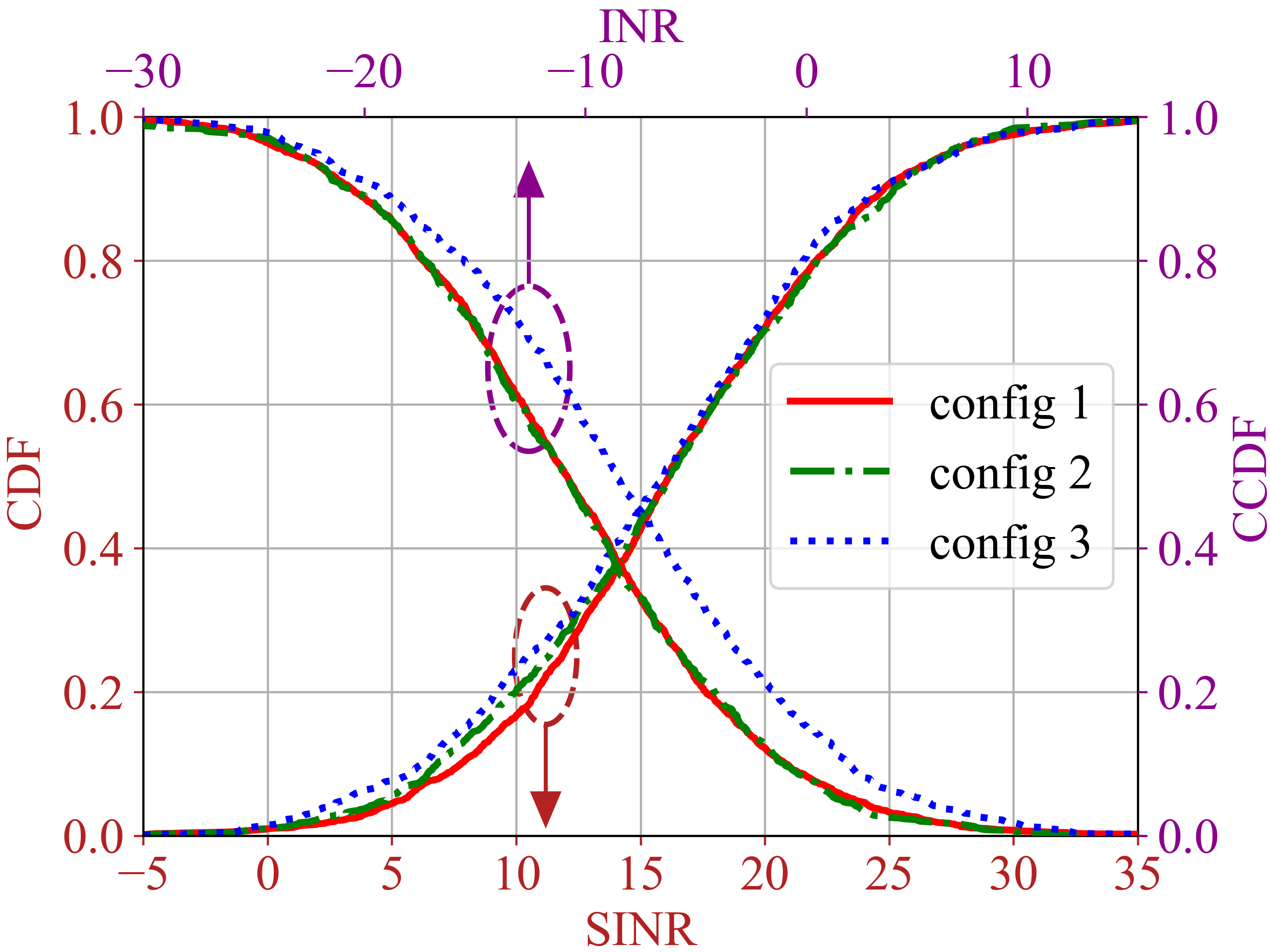}
 \caption{SINR and INR distributions of UEs for the different UAV penetration rates and power control
 configurations defined in Table~\ref{tab:pwc_config}.}
 \label{fig:sinr_inr}
\end{figure}

\begin{figure}[b]
\centering
\includegraphics[width=\columnwidth]{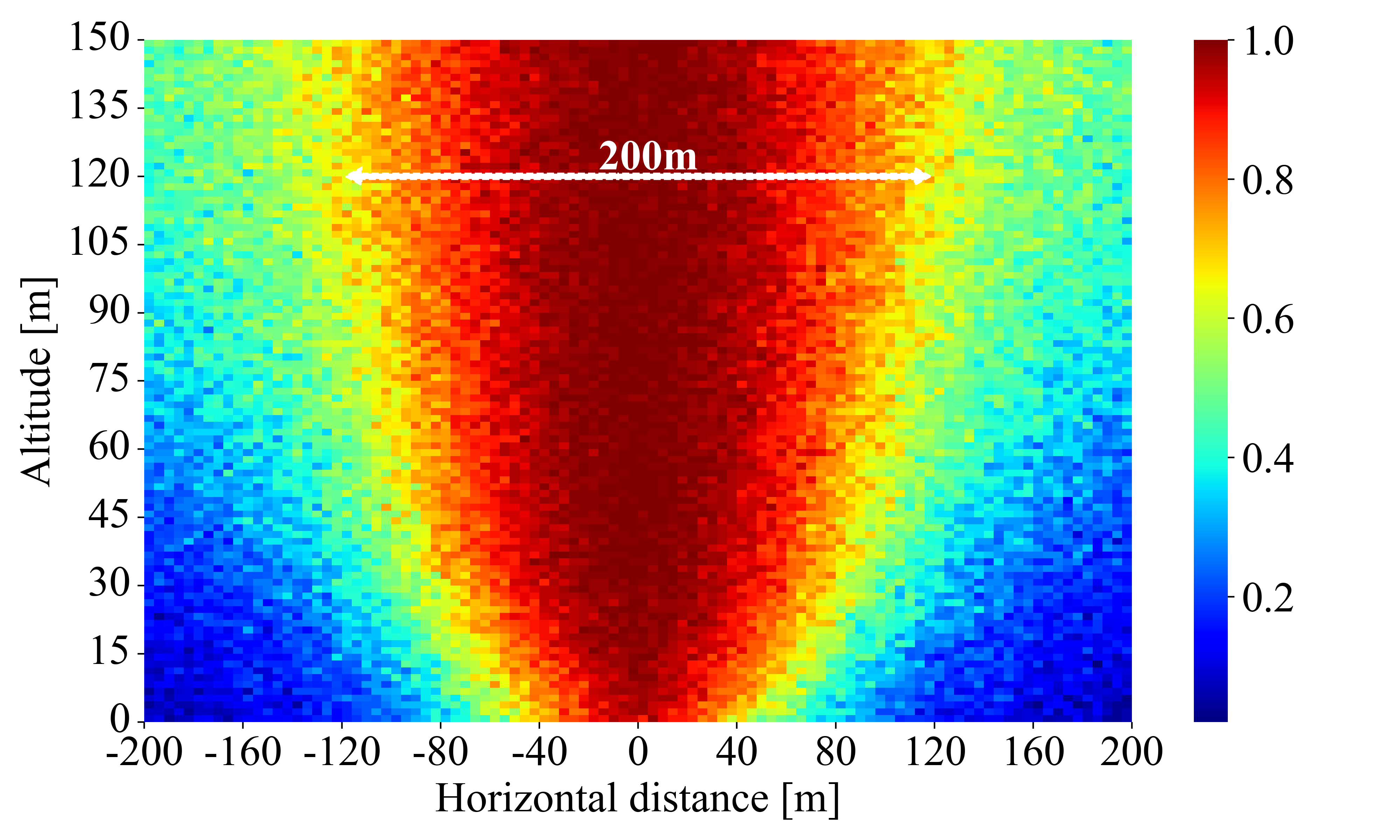}
\caption{LOS probability from a standard BS positioned at (0,0).}
\label{fig:los_prob}
\end{figure}

Fig.~\ref{fig:sinr_inr} shows the distribution of SINR and interference-to-noise ratio (INR) for UEs. Under SU-MIMO, the interference observed is only of intercell origin. The low INR values suggest that the network is predominantly noise-limited and that, as a consequence,
UAV power control improves the SINR of UEs by only 1-2 dB relative to full-power transmission (\emph{config 2} vs. \emph{config 3}). 
The reasons why the interference generated by UAVs onto UEs is rather negligible at mmWave frequencies are as follows:
\begin{itemize}
    \item LOS probability in urban scenarios increases only marginally with the UAV altitude. Fig.~\ref{fig:los_prob} shows the LOS probability of the aerial channel model developed in \cite{XiaRanMez2020}. Since UAVs are surrounded by buildings, the BS is highly visible by UAVs flying at $120$ m only within a horizontal distance of $200$ m. Hence, the chance to cause severe interference to UEs is low.
    \item The arrays of standard BSs are downtilted by $-12^\circ$, mitigating the interference from UAVs due to the reduced antenna element gains.
    \item Lastly, the increased pathloss and highly directional transmissions at mmWave frequencies result in a considerable attenuation of the interference. 
\end{itemize}

\subsection{MU-MIMO mmWave Communication}

\begin{figure}[!t]
  \centering
  \begin{minipage}[t]{0.95\columnwidth}
   \subfloat[UAVs]
        {{\includegraphics[width=1.0\linewidth]{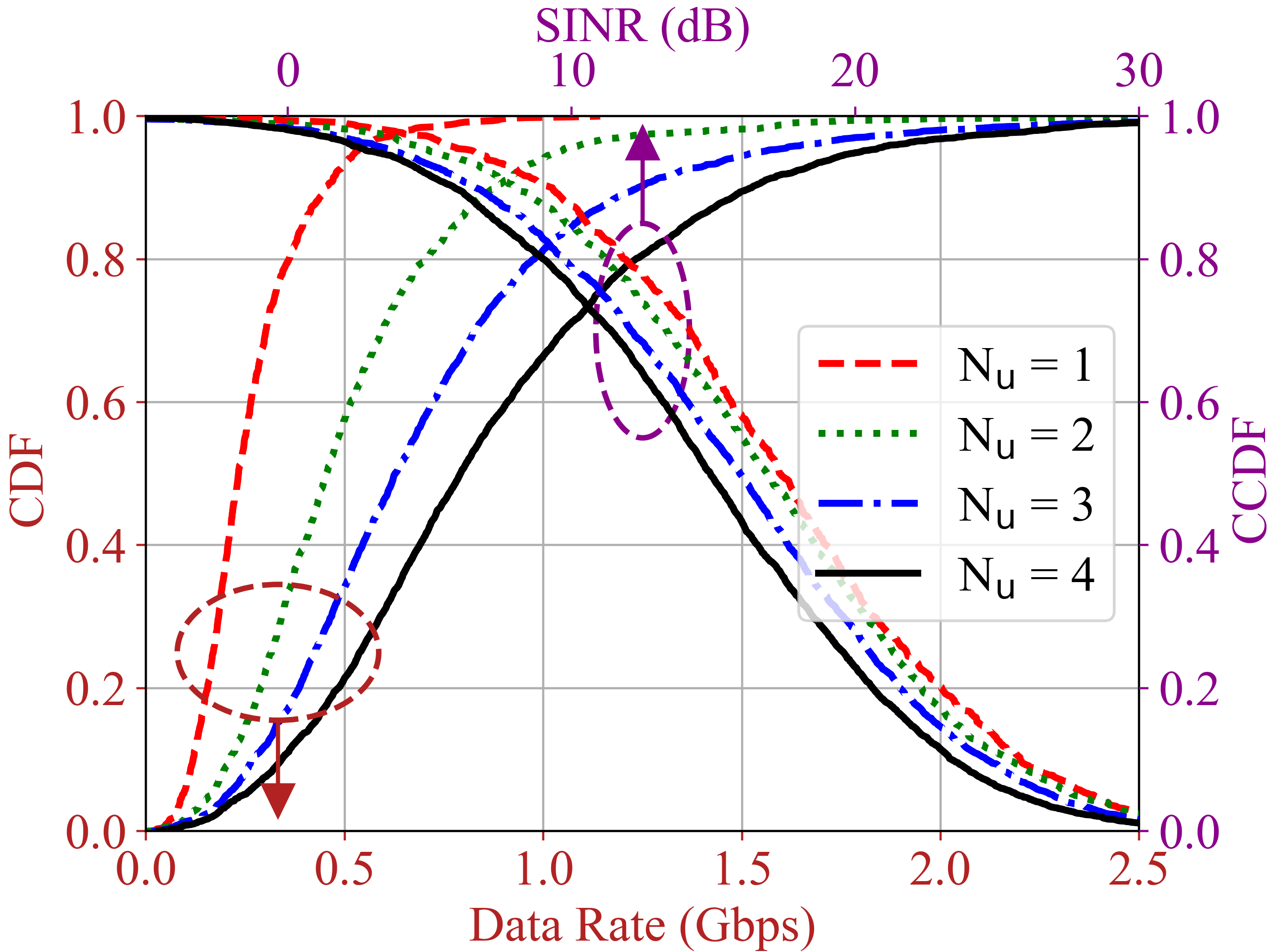}}}
        \quad
       \subfloat[UEs]
        {{\includegraphics[width=1.0\linewidth]{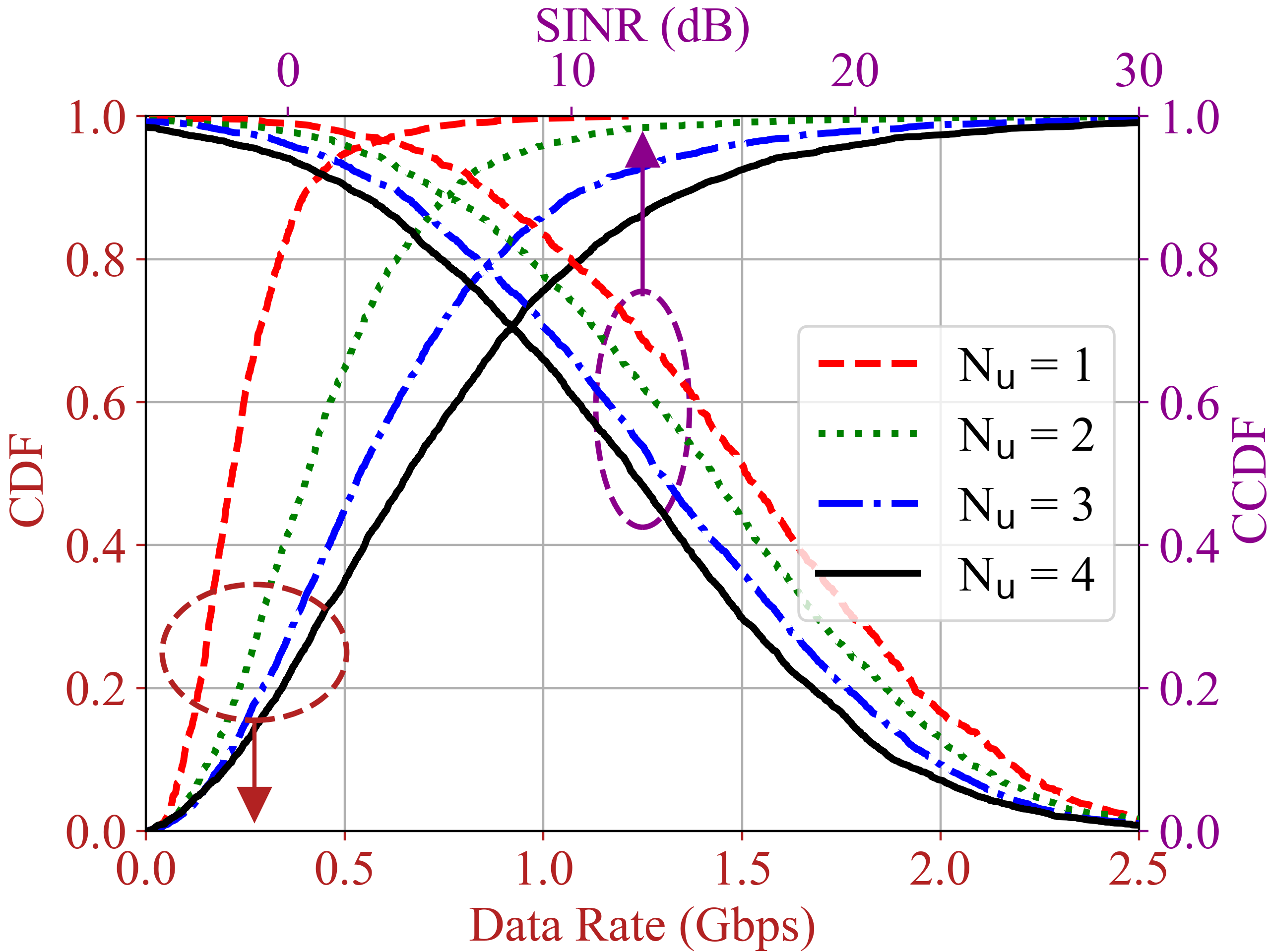}}}
      \caption{Performance under MU-MIMO with standard BSs, with $N_{\rm u}=1$ corresponding to SU-MIMO.}
      \label{fig:sdma}
  \end{minipage}
  \end{figure}

There is a limited amount of prior work on exploiting MU-MIMO in mmWave networks \cite{cuvelier2018mmwave, stirling2015multi, gomez2019optimal}. In particular, there is no established research on spectrum sharing between UEs and UAVs with MU-MIMO. 
 In this section, we tackle this issue by presenting the distribution of achievable data rates with an increasing number of simultaneous scheduled users.
 We consider $N_{\rm c}$ \emph{associated} users per cell, connected through the control channel but not necessarily scheduled in a particular 
 time slot.  In each time slot, the BS
 schedules $N_{\rm u}$ users and hence 
 each user is scheduled in a fraction 
 $N_{\rm u}/N_{\rm c}$ of the time slots.
 We simulated a scenario with $5$ UAVs and $5$ UEs per BS, so the average value of $N_{\rm c}=10$.  When scheduled, all users obtain
 access to the full bandwidth $B$ resulting
 in an average rate
%
%
\begin{equation}
  R = \frac{N_{\rm u}}{N_{\rm c}} B \log_2 ({1 + \mathrm{SINR}}).
\end{equation}
Fig.~\ref{fig:sdma} shows the data rate and SINR distributions of UAVs and terrestrial UEs with increasing active users. Due to MU-MIMO, the achievable data rate of both UAVs and UEs improves substantially. Although the number of scheduled users increases, the interference is still minimal. As shown in Fig.~\ref{fig:sdma},  the SINR coverage for UEs and UAVs is still ensured with a higher number of scheduled users ($N_{\rm u} = 4$).

\vspace{4mm}

%% file: 04_dedicated_cells.tex
\section{UAV-UE Spectrum Sharing with\\Dedicated Aerial mmWave Cells}
\label{sec:dedicated}

\input{0X_figure_spectrum_sharing}

In this section, we consider a hypothetical setup with MNOs operating in the same frequency band, namely:
\begin{itemize}
    \item A terrestrial operator, MNO$_{\textrm{T}}$, running a standard mmWave network as described in Section~\ref{sec:standard}.
    \item An aerial operator, MNO$_{\textrm{A}}$, running a dedicated mmWave network consisting of rooftop-mounted, uptilted BSs that are reserved exclusively for UAV communication.
\end{itemize}

As the penetration of UAVs increases, terrestrial MNO$_{\textrm{T}}$ may choose to share---under some leasing agreement---their spectrum with another MNO$_{\textrm{A}}$ that only intends to provide aerial connectivity, giving rise to the above multi-MNO scenario \cite{GerLopBen2022}. 
For this scenario, we examine the performance of both UAVs and UEs under two spectrum sharing paradigms: 
\begin{itemize}
\item \emph{Closed access},
where UAVs are only allowed to connect to dedicated BSs. On one hand, this paradigm requires low-to-no coordination for radio resource allocation, since all scheduling, beamforming,
and networking decisions are performed individually by each MNO. On the other hand, such simplification comes at the cost of a suboptimal spectrum usage, owing to the restricted UAV association and lack of synchronization.
\item \emph{Open access},
with UAVs allowed to connect to whichever BS offers the best quality of service, standard or dedicated. This enables a more efficient use of spectrum and
network infrastructure. However, a high coordination between the MNOs is required to jointly manage radio resources for aerial and terrestrial users, possibly entailing MNO$_{\textrm{T}}$ and MNO$_{\textrm{A}}$ to belong to the same network provider.
\end{itemize}

Fig.~\ref{fig:sharing} further summarizes the main features of the two multi-MNO spectrum sharing paradigms as compared to a single-MNO scenario. In the remainder of this section, we present the performance results obtained under the closed access and open access paradigms. We consider different dedicated BS densities by varying ISD$_{\rm d}$, and consider MU-MIMO with $N_{\rm u} = 2$. 

\begin{figure*}[!t]
  \centering
  \begin{minipage}[t]{.48\textwidth}
      \subfloat[UAVs]
          {{\includegraphics[width=1.0\linewidth]{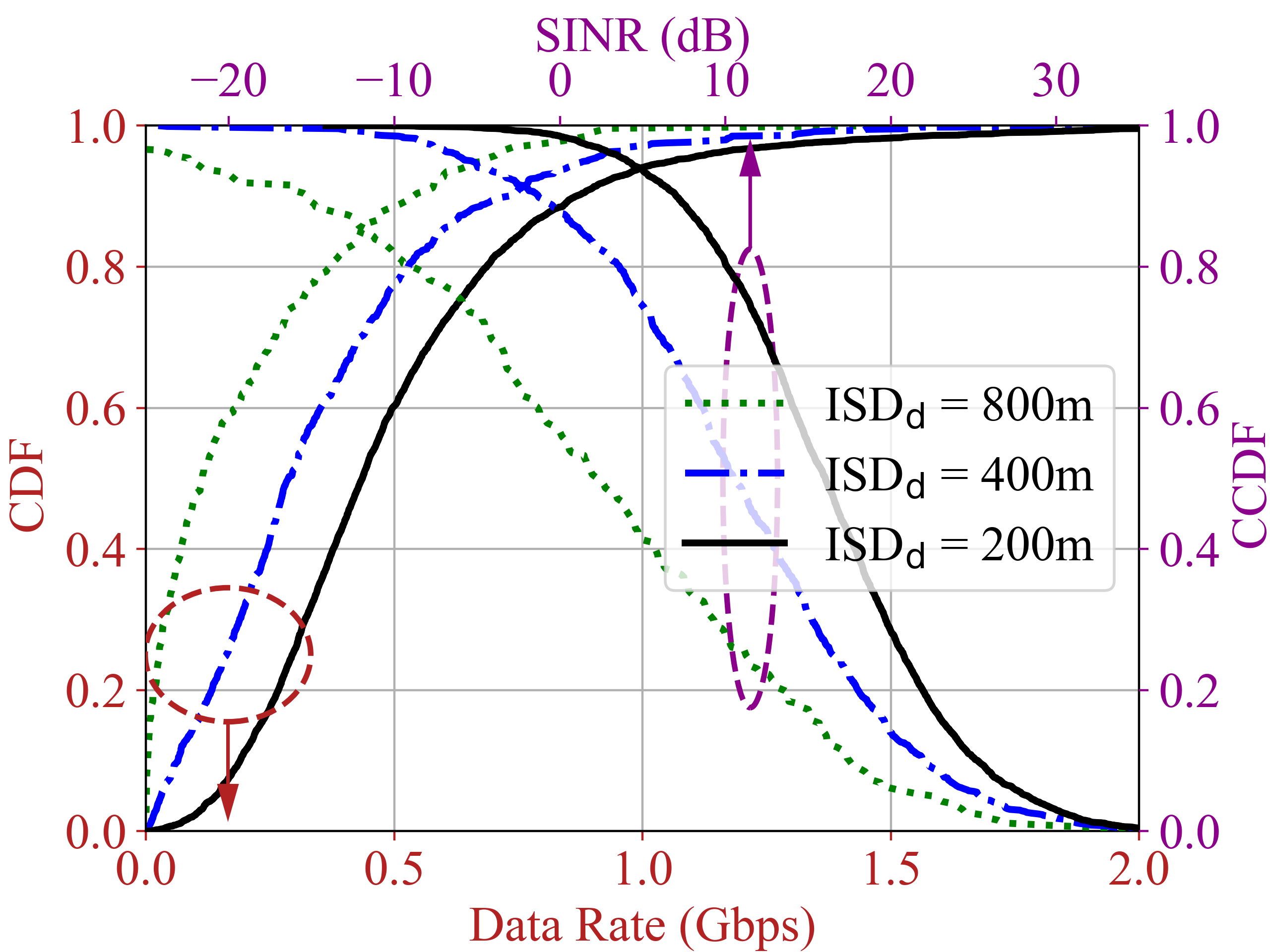}}}
        \quad
      \vspace{3mm}
      \subfloat[UEs]
          {{\includegraphics[width=1.0\linewidth]{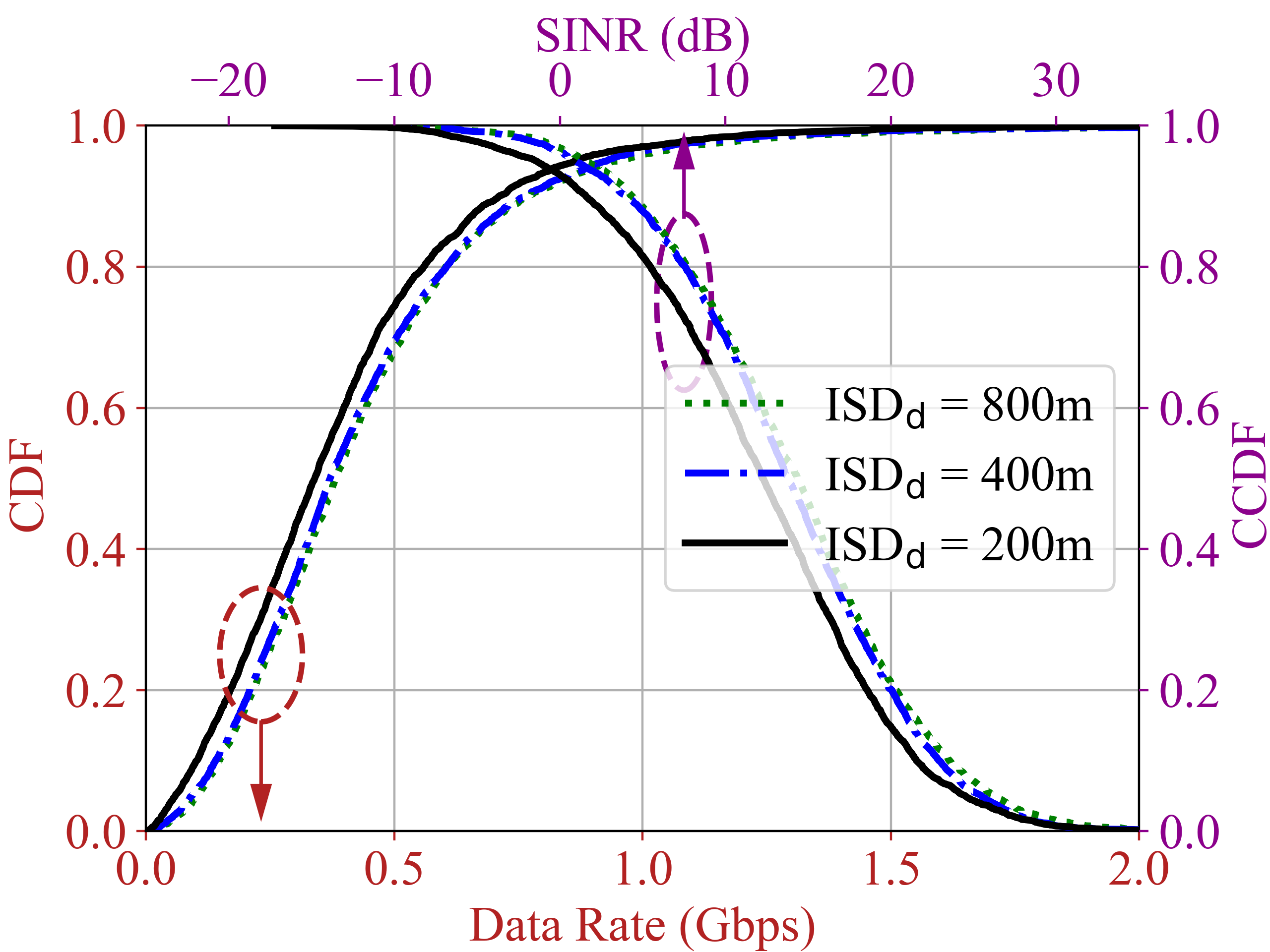}}}
        \caption{SINR and data rates for (a) UAVs and (b) terrestrial UEs under a Closed Access multi-operator paradigm. Both BS tiers feature MU-MIMO with $N_{\rm u}=2$.}
        \label{fig:inter}
  \end{minipage}
  \hfill
  \begin{minipage}[t]{.48\textwidth}
      \subfloat[UAVs]
          {{\includegraphics[width=1.0\linewidth]{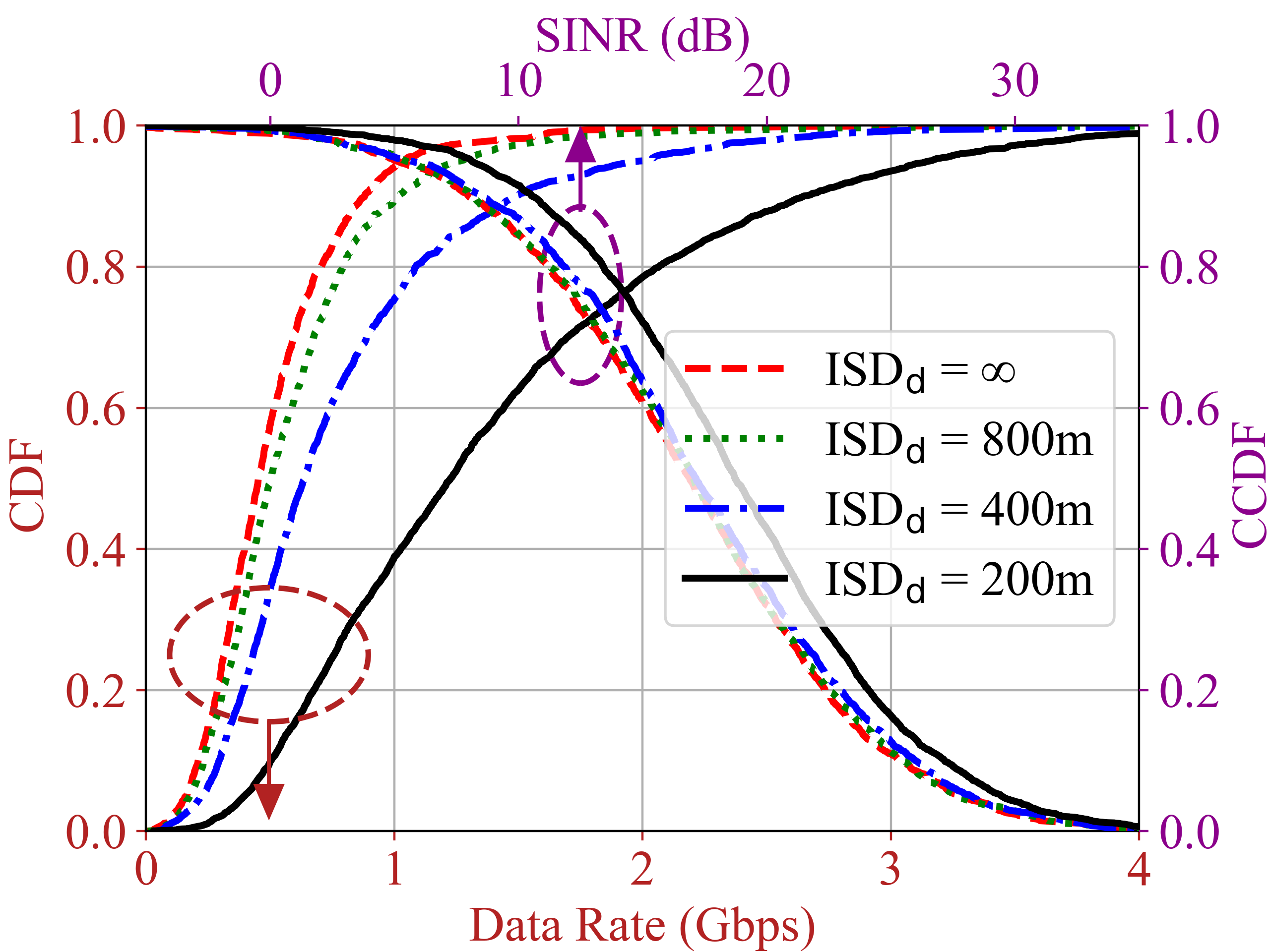}}}
     \quad
      \vspace{3mm}
      \subfloat[UEs]
          {{\includegraphics[width=1.0\linewidth]{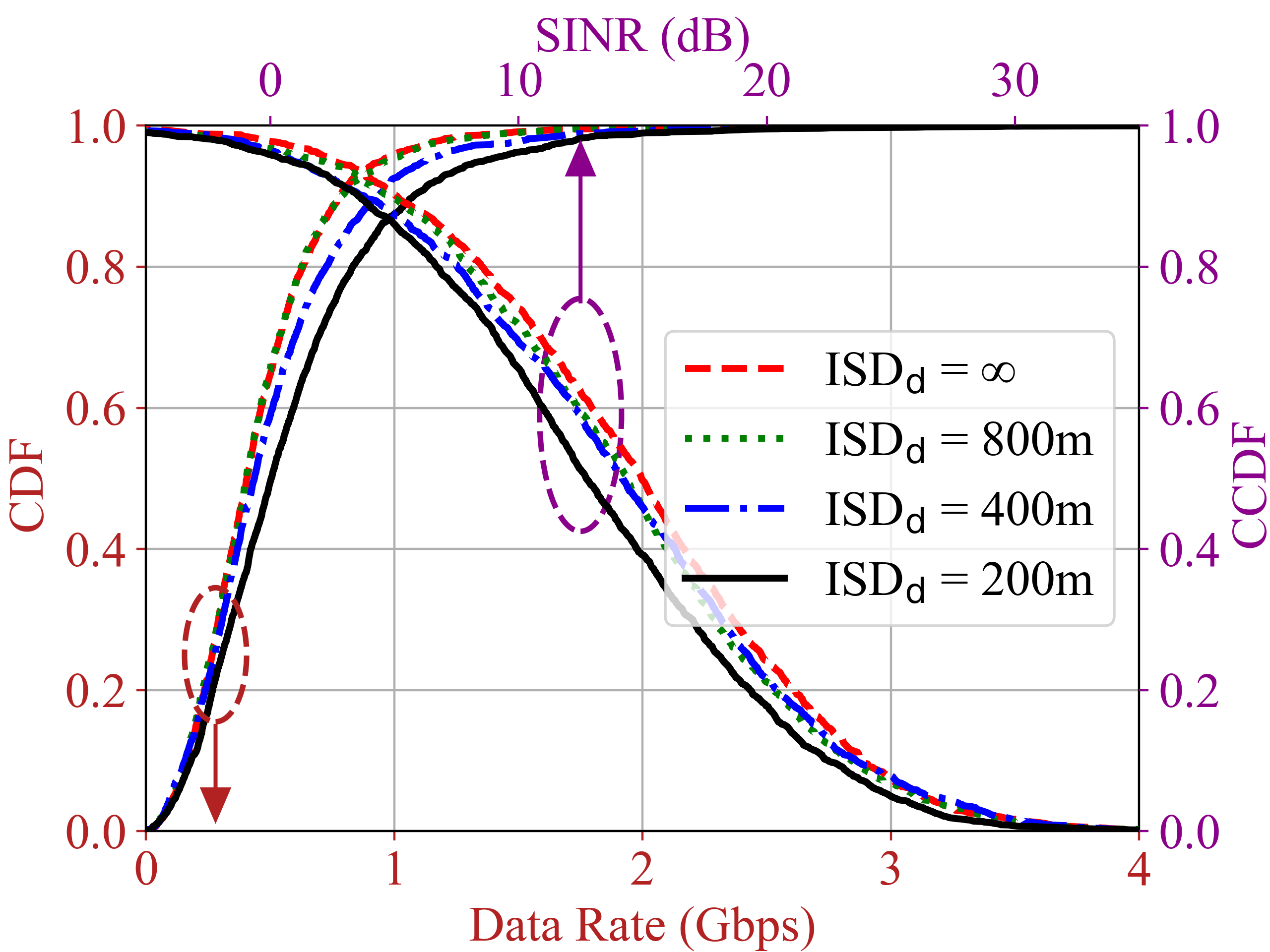}}}
            \caption{SINR and data rates for (a) UAVs and (b) terrestrial UEs under an Open Access multi-operator paradigm. Both BS tiers feature MU-MIMO with $N_{\rm u}=2$.}
        \label{fig:intra}
      \end{minipage}
  \end{figure*}

Fig.~\ref{fig:inter} shows the SINRs and data rates attained by UAVs and UEs under closed access, for various values of the dedicated BSs ISD. Since UAVs can only connect to dedicated BSs, the density of the latter greatly affects UAV coverage, with an ISD of 200~m or less required to keep the UAV outage at bay. As the density of dedicated BSs decreases, so does the number of concurrent UAV interferers seen by each standard BS. However, sparser dedicated deployments occasionally force UAVs to connect to far-flung dedicated BSs, thereby increasing their transmission power and generating stronger interference to standard BSs, an instance of the near-far problem. Overall, these two phenomena largely cancel one another, and the data rate experienced by UEs only marginally shrinks with the dedicated BS density.

Fig.~\ref{fig:intra} shows the SINRs and data rates achieved by UAVs and UEs under open access, with UAVs allowed to connect to the BS providing the highest signal strength, thereby circumventing the near-far effect. The figure considers three values for the dedicated BSs ISD, as well as the baseline case without dedicated deployment, labeled as ISD$_{\textrm{d}}=\infty$. Despite introducing additional interference on the UEs---see SINR curves in Fig.~\ref{fig:intra}(b)---, an increase in the number of dedicated BSs ultimately improves the data rates for both UEs and UAVs. Indeed, UEs can now access resources that were previously taken up by UAVs, as many of the latter are offloaded to dedicated BSs. In turn, UAVs achieve higher data rates because, when connecting to dedicated BSs, the link distance is reduced, uptilted arrays provide higher element gains, and resources are not shared with UEs.

%% file: 0X_figure_spectrum_sharing.tex
\begin{figure*}[t!]
   \centering
   \footnotesize
  \begin{tabular}{m{0.13 \textwidth}|m{0.24 \textwidth}|m{0.24\textwidth}|m{0.24\textwidth}|}
    \cline{2-4}
    & \centering \textbf{Single MNO}
    & \centering \textbf{Multi MNO --- Closed Access}
    & \centering \textbf{Multi MNO --- Open Access} \tabularnewline \cline{2-4}\hline
    \multicolumn{1}{|m{0.13\textwidth}|}
        {\centering \textbf{Spectrum reuse and efficiency}} 
        &\cellcolor{red!25}  \textbf{Low:}  
        No dedicated deployment; no spectrum reuse between aerial and terrestrial networks.
        & \cellcolor{orange!25}  \textbf{Medium:} Suboptimal spectrum and infrastructure reuse due to restricted association
        and lack of synchronization.
        & \cellcolor{green!25} \textbf{High:} 
        Optimal spectrum and infrastructure reuse;
        seamless interference and user association 
        coordination. 
        \tabularnewline \hline\hline
    \multicolumn{1}{|m{0.13\textwidth}|}
    {\centering \textbf{Coordination for resource allocation}}
    & \cellcolor{red!25}  \textbf{None:}  
    Single operator running autonomously, no need for coordination.
    & \cellcolor{orange!25} \textbf{Low:} 
    Requires monitoring inter-operator interference but scheduling, beamforming, and networking decisions are made individually.
    & \cellcolor{green!25} \textbf{High:} 
    Requires operators to manage 
    aerial and terrestrial users jointly.
    \tabularnewline \hline            
      
     \end{tabular}

    \caption{Spectrum sharing paradigms: (left) single MNO, (center) two MNOs in closed access, and (right) two MNOs in open access.}
  
  \label{fig:sharing}
\end{figure*}

%% file: 05_conclusion.tex
\section{Conclusion}

The goal of this paper was to understand the effect of mmWave spectrum sharing among UAVs and terrestrial users. To this end, an extensive campaign of simulations was launched, incorporating an accurate directional mmWave aerial channel model and UAV antenna patterns alongside 3GPP-based terrestrial mmWave channels and BS antenna patterns. 

We considered multiple scenarios: (i) one with a single terrestrial operator running a standard mmWave network with SU-MIMO and MU-MIMO, (ii) another where a second aerial operator reuses the same spectrum independently to support UAVs via dedicated uptilted cells, (iii) and one last scenario where the two operators jointly offer connectivity to UAVs. Multiple insights have been drawn:
\begin{itemize}
\item The interference generated by UAVs onto terrestrial UEs in standard mmWave networks is minimal, thanks to downtilted cells and high directionality.
\item Owing to the above intrinsic spatial separation, MU-MIMO is highly effective and data rates improve despite the additional concurrent interferers.
\item Co-channel uptilted BSs run independently by an aerial operator in closed access mode can provide satisfactory UAV coverage, provided they are densely deployed.
\item In open access---where UAVs can connect to either standard or dedicated uptilted cells---such uptilted cells are no longer crucial for UAV coverage, but their addition boosts the data rates of UAVs and UEs alike.
\end{itemize}
